\documentclass{jetpl}
\twocolumn

\lat

\title{Search for high-energy neutrinos from GW170817 with the Baikal-GVD
         neutrino telescope}

\rtitle{Search for high-energy neutrinos\ldots}

\sodtitle{Search for high-energy neutrinos from GW170817 with Baikal-GVD
         neutrino telescope}

\author{A.\,D.\,Avrorin$^{1}$,
A.\,V.\,Avrorin$^{1}$,
V.\,M.\,Aynutdinov$^{1}$,
R.\,Bannash$^{7}$,
I.\,A.\,Belolaptikov$^{2}$,
V.\,M.\,Brudanin$^{2}$,
N.\,M.\,Budnev$^{3}$,
A.\,A.\,Doroshenko$^{1}$,
G.\,V.\,Domogatsky$^{1}$,
R.\,Dvornický$^{2,8}$,
A.\,N.\,Dyachok$^{3}$,
Zh.-A.\,M.\,Dzhilkibaev$^{1+}$\/\thanks{e-mail: djilkib@yandex.ru},
L.\,Fajt$^{2,8,9}$,
S.\,V.\,Fialkovsky$^{5}$,
A.\,R.\,Gafarov$^{3}$,
K.\,V.\,Golubkov$^{1}$,
T.\,I.\,Gres$^{3}$,
Z.\,Honz$^{2}$,
K.\,G.\,Kebkal$^{7}$,
O.\,G.\,Kebkal$^{7}$,
E.\,V.\,Khramov$^{2}$,
M.\,M.\,Kolbin$^{2}$,
K.\,V.\,Konischev$^{2}$,
A.\,P.\,Korobchenko$^{2}$,
A.\,P.\,Koshechkin$^{1}$,
V.\,A.\,Kozhin$^{4}$,
V.\,F.\,Kulepov$^{5}$,
D.\,A.\,Kuleshov$^{1}$,
M.\,B.\,Milenin$^{5}$,
R.\,A.\,Mirgazov$^{3}$,
E.\,R.\,Osipova$^{4}$,
A.\,I.\,Panfilov$^{1}$,
L.\,V.\,Pan'kov$^{3}$,
D.\,P.\,Petukhov$^{1}$,
E.\,N.\,Pliskovsky$^{2}$,
M.\,I.\,Rozanov$^{6}$,
E.\,V.\,Rjabov$^{3}$,
V.\,D.\,Rushay$^{2}$,
G.\,B.\,Safronov$^{2}$,
F.\,Simkovic$^{2,8}$,
A.\,V.\,Skurikhin$^{4}$,
B.\,A.\,Shoibonov$^{2}$,
A.\,G.\,Solovjev$^{2}$,
M.\,N.\,Sorokovikov$^{2}$,
M.\,D.\,Shelepov$^{1}$,
I.\,Shtekl$^{2,9}$,
O.\,V.\,Suvorova$^{1}$,
V.\,A.\,Tabolenko$^{3}$,
B.\,A.\,Tarashansky$^{3}$,
S.\,A.\,Yakovlev$^{7}$,
A.\,V.\,Zagorodnikov$^{3}$,
V.\,L.\,Zurbanov$^{3}$}

\rauthor{A.\,D.\,Avrorin A.D., A.\,V.\,Avrorin A.V.\ldots}

\sodauthor{Avrorin, Avrorin}

\address{$^1$Institute for Nuclear Research RAS,
117312 Moscow, Russia\\~\\
$^2$Joint INstitute for Nuclear Research, 141980 Dubna, Russia\\~\\
$^3$Irkutsk State University, 664003 Irkutsk, Russia\\~\\
$^4$Institute of Nuclear Physics, Moscow State University, 119991 Moscow, Russia\\~\\
$^5$Nizhni Novgorod State Technical University, 603950 Nizhni Novgorod, Russia\\~\\
$^6$St. Petersburg State Marine Technical University, 190008 St. Petersburg, Russia\\~\\
$^7$EvoLogics, 13355 Berlin, Germany\\~\\
$^8$Comenius University, 84248 Bratislava, Slovakia\\~\\
$^9$Czech Technical University, 12800 Prague, Czech Republic}

\dates{24 October 2018}{*}

\abstract{The Advanced LIGO and Advanced Virgo observatories recently
discovered gravitational waves from a binary neutron star
inspiral. A short gamma-ray burst (GRB) that followed the merger of
this binary was also recorded by Fermi-GBM and INTEGRAL,
indicating particle acceleration by the source.
The precise location of the event was determined by
optical detections of emission following the merger. We searched for
high-energy neutrinos from the merger in the TeV - 100 PeV
energy range using Baikal-GVD. No neutrinos directionally coincident
with the source were detected within $\pm$500 s around the merger time,
as well as during a 14-day period after the GW detection.
We derived 90\% confidence level upper limits on the neutrino fluence
from GW170817 during a $\pm$500 s window centered on the GW trigger
time, and a 14-day window following the GW signal under the assumption
of an $E^{-2}$ neutrino energy spectrum.}


\begin{document}

\maketitle

{\bf Introduction.}
 A gravitational wave signal, GW170817, from a
binary neutron star merger has been recorded
by the Advanced LIGO and Advanced Virgo
observatories on August 17, 2017 \cite{GW}.
A short GRB (GRB170817A), associated with GW170817, was detected
by Fermi-GBM and INTEGRAL. NGC 4993 was localized as the host
galaxy of the merger by follow up optical observations.
High-energy neutrino signals
associated with the merger were searched for by the ANTARES
and IceCube neutrino telescopes in muon and cascade modes
and the Pierre Auger Observatory \cite{GWNU_1} and
Super-Kamiokande \cite{GWNU_2}. Two different time windows
were used for the searches. First, a $\pm$500 s time
window around the merger was used to search for neutrinos associated
with prompt and extended gamma-ray emission \cite{Baret,Kimura}.
Second, a 14-day time window following the GW detection, to cover
predictions of longer-lived emission processes \cite{Gao,Fang}.
No significant neutrino signal
was observed by the neutrino telescopes.

%%%%%%%%%%%%%%%%%%%%%%%%%%%%%%%%%%%%%%%%%%%%%%%%%%%%%%%%%%%%%

The deep underwater neutrino telescope Baikal Gigaton Volume Detector
(Baikal-GVD) is currently under construction in Lake Baikal \cite{RefJ}.
The telescope has a modular structure
and consists of functionally independent sub-arrays (clusters)
of optical modules (OMs). Since each GVD-cluster represents a
multi-megaton scale Cherenkov detector, studies of neutrinos of
different origin are allowed at early stages of construction.
During 2017 two GVD-clusters have been operated in Lake Baikal.
In this Letter we present searches for high-energy neutrinos in
coincidence with GW170817/GRB170817A by the Baikal-GVD high-energy
neutrino telescope.

%%%%%%%%%%%%%%%%%%%%%%%%%%%%%%%%%%%%%
\begin{figure}
\includegraphics[width=0.45\textwidth,height=7cm]{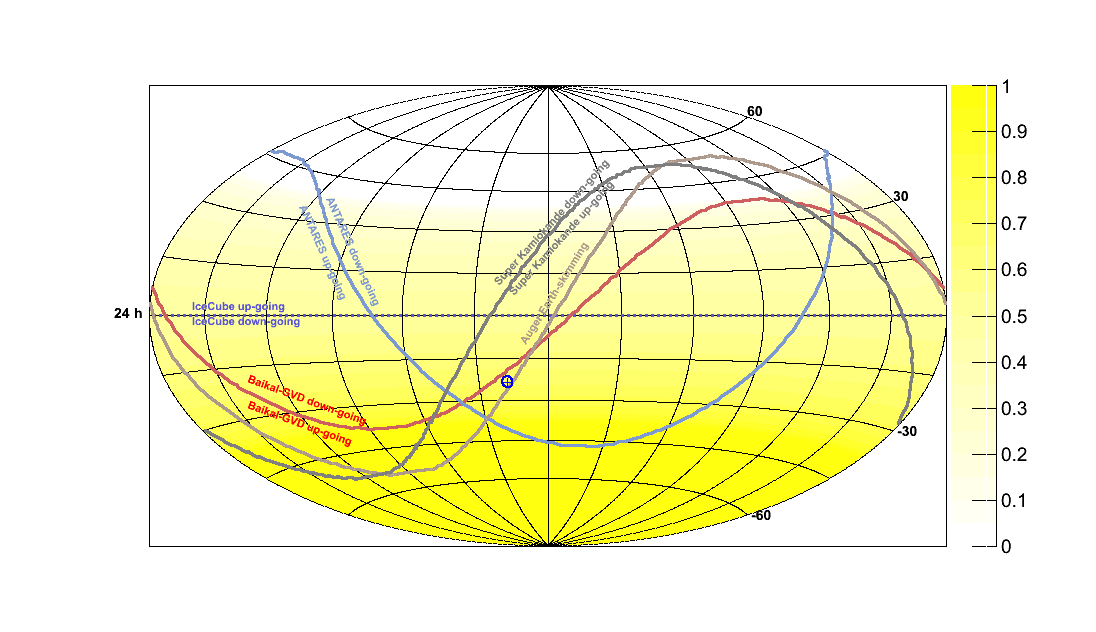}
\caption{{\bf Figure 1.}
Localizations of NGC 4993 and horizons separating down-going
and up-going neutrino directions
for IceCube, ANTARES, SuperKamiokande and Baikal-GVD at the
time of the GW event in equatorial coordinates.
The zenith angle of the source at the
detection time of the merger was 73.8$^{\circ}$ for ANTARES,
66.6$^{\circ}$ for IceCube,
108$^{\circ}$ for SK and 93.3$^{\circ}$ for Baikal-GVD.
}
\label{fig.1.1}
\end{figure}
%%%%%%%%%%%%%%%%%%%%%%%%%%%%%%%%%%%%%%%%%%%%%%%%%%

{\bf Detector and search method.}
The geographical coordinates of the Baikal-GVD site
are 51$^{\circ}$50$^{\prime}$N and 104$^{\circ}$20$^{\prime}$E.
The detector instruments the deep water of Lake Baikal
with optical modules - pressure resistant glass spheres
equipped with photomultiplier
tubes (PMT) Hamamatsu R7081-100 with photocathode diameter of
10” and  a quantum efficiency of $\sim$35\% \cite{OM}. The PMTs record the
Cherenkov radiation from secondary particles produced in
interactions of high-energy neutrinos inside or near the
instrumented volume. From the arrival times of light at the
PMTs and from the amount of light, direction and energy of the
incoming neutrinos are derived. Baikal-GVD in it's 2017 design
consisted of two clusters - each of them with 288 optical modules.
A cluster
comprises eight vertical strings attached to the lake floor:
seven side strings on a radius of 60 m around a central one.
Each string carries 36 OMs, arranged at depths between 735
and 1260 meters (525 m instrumented length). The vertical
spacing between the OMs along a string is 15 m. The OMs on each
string are functionally combined in 3 sections. A section comprises
12 OMs with data processing and communication electronics and
forms a detection unit (DU) of the array. All analogue
signals from the PMTs are digitized, processed in the
sections and sent to shore if certain trigger conditions
(e.g. a minimum number of fired PMTs) are fulfiled \cite{DAQ}.

IceCube discovered a diffuse flux of high-energy
astrophysical neutrinos in 2013 \cite{IC1}.
The data sample of their high-energy starting event analysis
(HESE, 7.5 year sample) comprises 103 events, 77 of which are
identified as cascades and 26 as track events \cite{IC2}.
These results demonstrate the importance of the cascade
mode of neutrino detection with neutrino
telescopes.
A search for high-energy neutrinos associated  with GW170817 in Baika-GVD
is based on the selection of cascade events generated by neutrino
interactions in the sensitive volume of array.
The procedure for reconstructing the parameters of high-energy showers -
the shower energy, direction, and vertex - is performed in two steps.
In the first step, the shower vertex coordinates are reconstructed
by $\chi_t^2$ minimization using
the time information from the telescope's triggered photo-sensors.
The reconstruction quality can be increased by applying additional event
selection criteria based on the limitation of the admissible values for
the specially chosen parameters characterizing the events.
In the second step, the shower energy and direction are reconstructed by
applying the maximum-likelihood and using the shower coordinates
reconstructed in the first step.
The values of the variables $\theta$, $\phi$, and E$_{sh}$ corresponding
to the maximum value of the likelihood are chosen as the polar and
azimuth angles characterizing the direction and the
shower energy.
%%%%%%%%%%%%%%%%%%%%%%%%%%%%%%%%%%%%%
\begin{figure}
\includegraphics[width=0.45\textwidth]{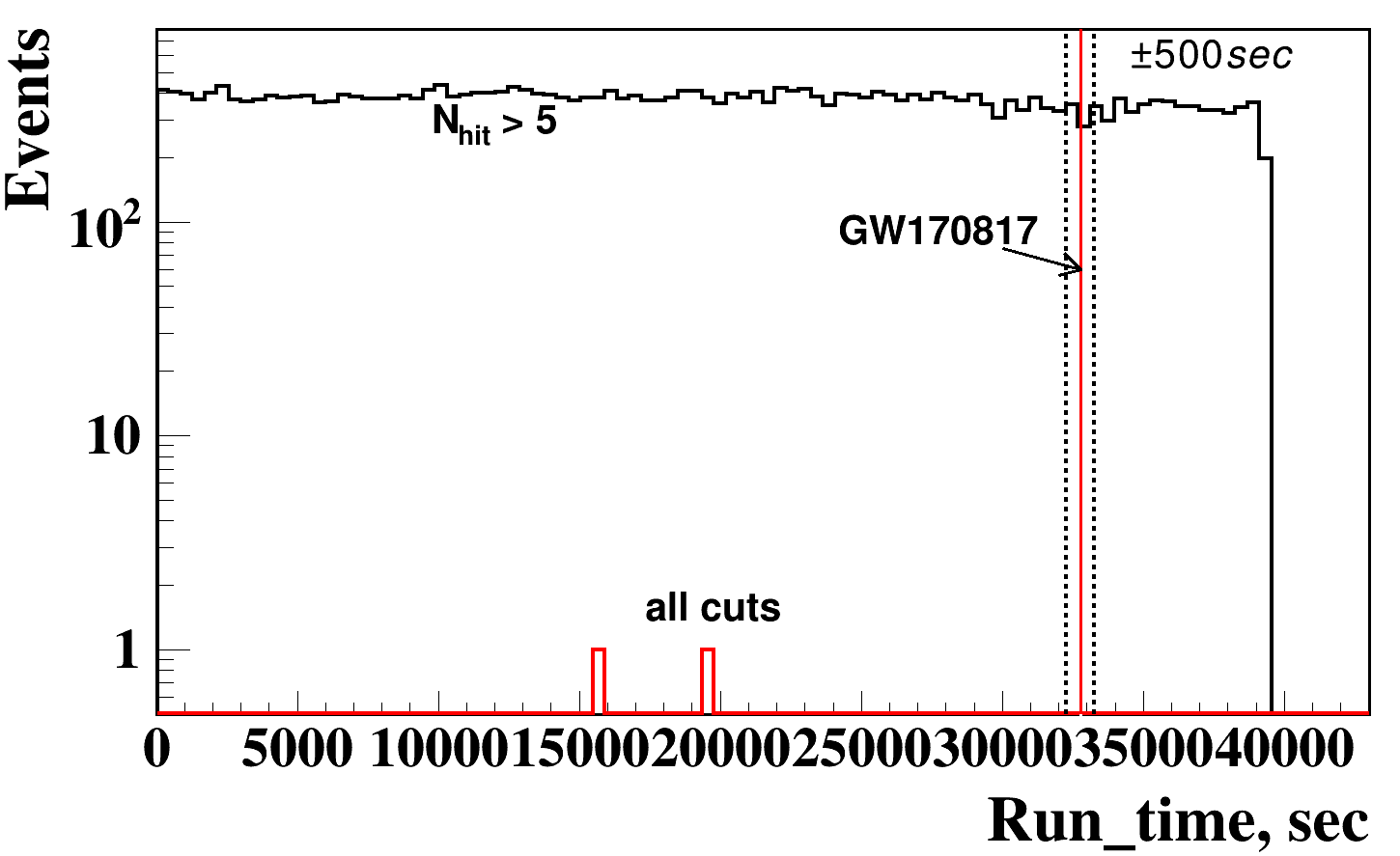}
\caption{{\bf Figure 2.}
Temporal distribution of events during the data taking run contaning
the $\pm$500 s time window around the GW event. The histograms represent
events with hit OMs N$_{hit}>$5, and events
surviving all selection cuts used for the neutrino search within $\pm$500 s
time window around the GW event.
}
\label{fig.1.2}
\end{figure}
%%%%%%%%%%%%%%%%%%%%%%%%%%%%%%%%%%%%%%%%%%%%%%%%%%

%%%%%%%%%%%%%%%%%%%%%%%%%%%%%%%%%%%%%
\begin{figure}[h]
\includegraphics[width=0.45\textwidth]{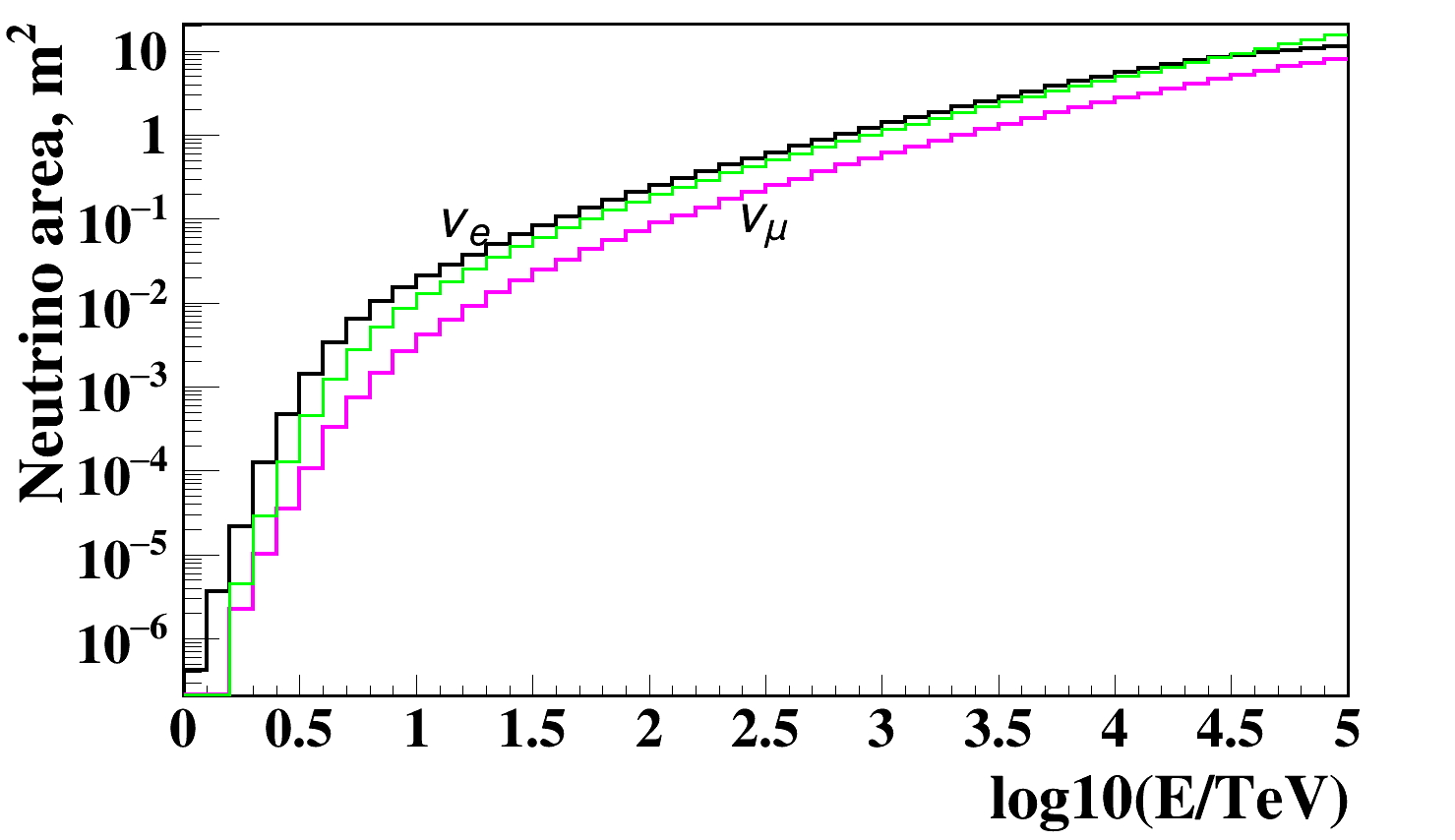}
\hfill
\includegraphics[width=0.46\textwidth]{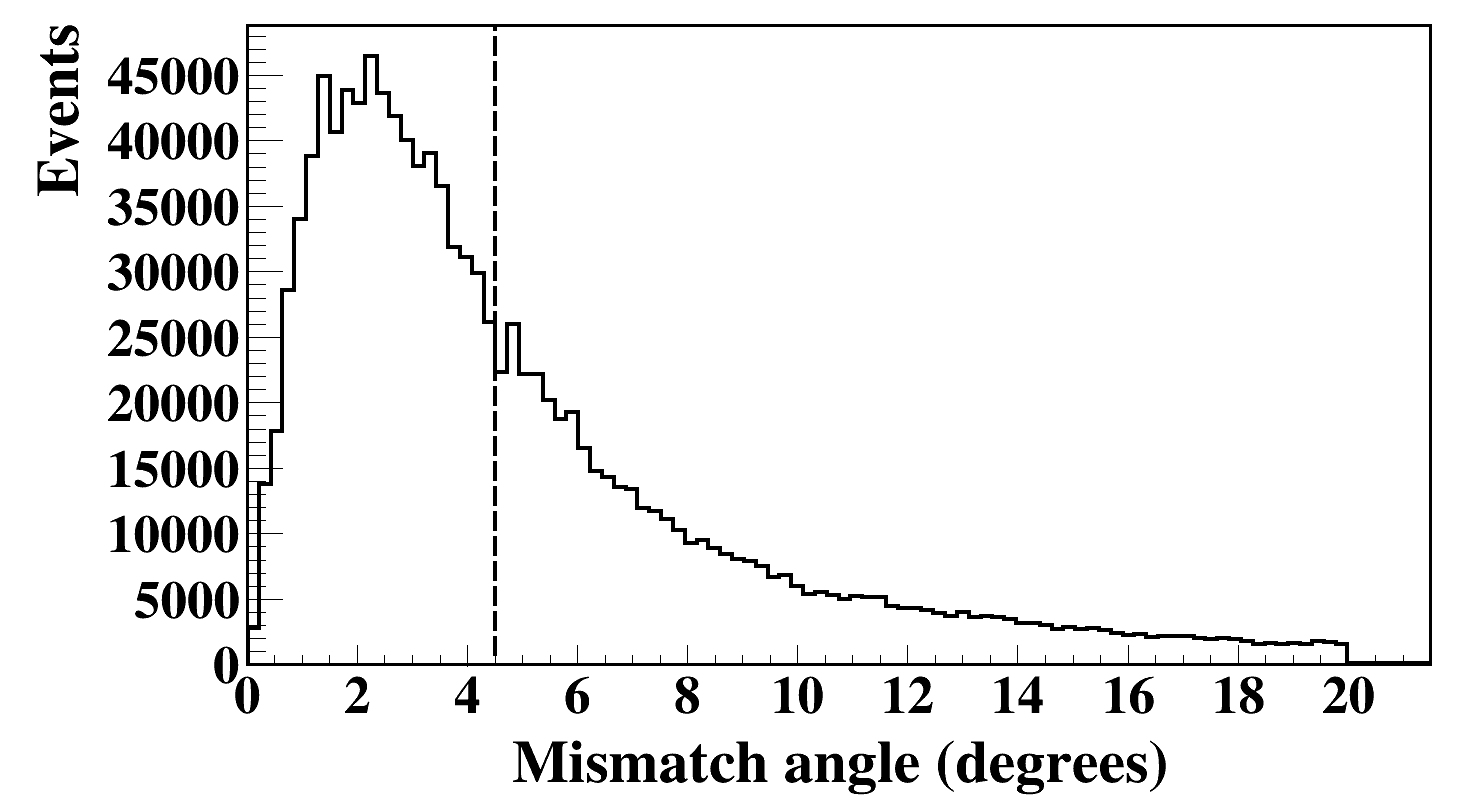}
\\
\caption{{\bf Figure 3.}
Neutrino effective areas (top panel)
and error angles distribution (bottom panel).
}
\label{fig3}
\end{figure}
%%%%%%%%%%%%%%%%%%%%%%%%%%%%%%%%%%%%%%%%%%%%%%%%%%

The zenith angle of NGC 4993 at the detection time
of GW170817 was 93.3$^{\circ}$ for Baikal-GVD
(see, Figure \ref{fig.1.1}). Since background
events from atmospheric muons and neutrinos can be significantly
suppressed by requiring time and space coincidence with the GW signal,
relatively weak cuts can be used for neutrino selection. For the search for
neutrino events within a $\pm$500 s window around the GW event, 731 events were
selected, which comprise >5 hit OMs at >2 hit strings. After applying
cascade reconstruction procedures and dedicated quality cuts, two events
were selected. Finally, requiring directional coincidence with NGC 4993
$\psi < 20^{\circ}$ no neutrino candidates survived.
Shown in Figure \ref{fig.1.2} are temporal distributions of events
fulfilling the initial selection requirement as well as
events surviving all cuts during the 39347 s long data
taking run, which contains the $\pm$500 s time window around GW170817.
Neutrino effective areas for each flavour are shown in Figure \ref{fig3}
(top panel). Error angles
distribution is shown in Figure \ref{fig3} (bottom panel).
The median angular error is
4.5$^{\circ}$ with this set of relaxed cuts and
the expected number of atmospheric background
events is about $5 \times 10^{-2}$ during the coincident time window.

\begin{figure}
\includegraphics[width=0.45\textwidth]{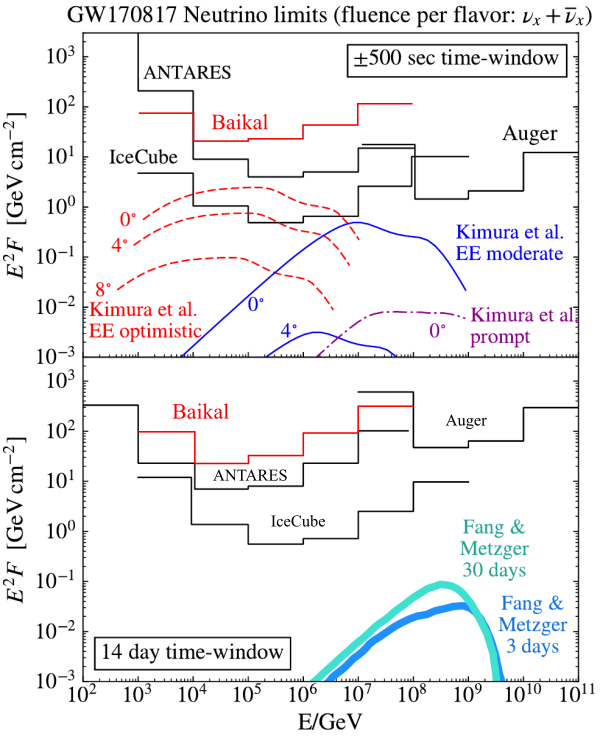}
\caption{{\bf Figure 4.}
Upper limits (at 90 \% confidence level) on the neutrino
spectral fluence from GW170817 during a $\pm500$ s window centered
on the GW trigger time (top panel), and a 14-day window following
the GW trigger (bottom panel). For each experiment, limits are
calculated separately for each energy decade, assuming a spectral
fluence $F(E) = F_{up} \times [E/GeV]^{-2}$ in that decade only. Also
shown are predictions by neutrino emission models (see \cite{GWNU_1}
for details).
%Also
%shown are predictions by neutrino emission models. In the upper
%plot, models from Kimura et al. (2017) for both extended emission
%(EE) and prompt emission are scaled to a distance of 40 Mpc, and
%shown for the case of on-axis viewing angle ($0^{\circ}$ ) and selected off-
%axis angles to indicate the dependence on this parameter. GW data
%and the redshift of the host-galaxy constrain the viewing angle to
%$\theta \in [0^{\circ} , 36^{\circ} ]$ (see Section 3).
%In the lower plot, models from Fang
%\& Metzger (2017) are scaled to a distance of 40 Mpc. All fluences
%are shown as the per flavor sum of neutrino and anti-neutrino fluence,
%assuming equal fluence in all flavors, as expected for standard
%neutrino oscillation parameters.
}
\label{fig4}
\end{figure}

The absence of neutrino candidates in the $\pm$500 s window associated with
GW170817 allows to constrain the fluence of neutrinos from GW170817A.
Assuming an $E^{-2}$ spectrum single-flavor differential limits to
the spectral fluence in bins of one decade in energy have been derived
according to \cite{Feld} 
(see Figure \ref{fig4} (top panel)).
In the range from 5 TeV to 10 PeV
a 90\% CL upper limit on an $E^{-2}$ power-law
spectral neutrino fluence is 5.2$\times(E/GeV)^{-2}$ GeV$^{-1}$cm$^{-2}$.

The search over 14 days used a more stringent cut on the number of
hit OMs - N$_{hit} >$7. The zenith angle of the optical counterpart
 oscillates daily between 74$^{\circ}$ and 150$^{\circ}$.
No events spatially coincident with GRB170817A were found in this search.
Given the non-detection
of neutrino events associated with GW170817, differential
upper limits have been derived (see Figure \ref{fig4}
(bottom panel)). The corresponding upper limit to the
spectral fluence is 9.0$\times(E/GeV)^{-2}$ GeV$^{-1}$cm$^{-2}$
over the same energy range as for the $\pm$500 s time window.

%%%%%%%%%%%%%%%%%%%%%%%%%%%%%%%%%%%%%
%%%%%%%%%%%%%%%%%%%%%%%%%%%%%%%%%%%%%%%%%%%%%%%%%%

{\bf Conclusion.}
The deep underwater neutrino telescope Baikal-GVD
is currently under construction.
During 2017 two GVD-clusters have been operated in Lake Baikal.
We performed a search for high-energy neutrinos associated with the
gravitational wave, GW170817, from the binary neutron star merger in
NGS4993 with Baikal-GVD, using the cascade detection mode. The search
was performed within a $\pm$500 s time window around GW170817,
and 14 days after detection of the gravitational wave. No neutrino
events associated with the merger were detected within the $\pm$500 s time interval,
nor during 14 days after the gravitational wave. This allows to derive the neutrino
spectral fluence limits from this merger. The ongoing analysis of GVD data using
the muon detection mode will allow to improve these limits.

{\it This work was supported by the Russian Foundation for Basic
Research (Grants 16-29-13032, 17-02-01237).}
%%%%%%%%%%%%%%%%%%%%%%%%%%%%%%%%%%%%%%%%%%%%%%%%%

\end{document}